\documentclass[12pt]{article} 

\pdfoutput=1

\usepackage[margin=25mm]{geometry}
\usepackage{amsmath,amsfonts,amssymb}
\usepackage{graphicx}
\usepackage{setspace}
\usepackage{tocloft}
\usepackage{bm}
\usepackage{siunitx}
\usepackage{float}
\usepackage{authblk}
\usepackage{cite}
\usepackage[format=hang]{caption}

\providecommand{\keywords}[1]
{
  \small	
  \textbf{\textit{Keywords---}} #1
}

\title{\textbf{Influence of an external electric field on the energy dissipation at the initial stage of laser ablation}}

\author[a]{Steffen Hirtle}
\author[a]{Pavel N. Terekhin}
\author[b]{Mareike Sch\"afer}
\author[b]{Yiyun Kang}
\author[a]{Sanjay Ashok}
\author[b]{Johannes A. L’huillier}
\author[a]{Baerbel Rethfeld}
\affil[a]{\small Department of Physics and Research Center OPTIMAS, Technische Universität Kaiserslautern, Erwin-Schrödinger-Straße 46, 67663 Kaiserslautern, Germany}
\affil[b]{\small Photonik-Zentrum Kaiserslautern e.V. and Research Center OPTIMAS of Technische Universität Kaiserslautern, Kohlenhofstraße 10, 67663 Kaiserslautern, Germany}

\cftpagenumbersoff{figure}
\cftpagenumbersoff{table} 
\begin{document} 
\maketitle

\begin{abstract}

\noindent A density-dependent two-temperature model is applied to describe laser excitation and the following relaxation processes of silicon in an external electric field. Two approaches on how to describe the effects of the external electric field are presented. The first approach avoids the buildup of internal electric fields due to charge separation by assuming ambipolar diffusion and adds an additional carrier-pair current. In the second approach, electrons and holes are treated separately to account for charge separation and the resulting shielding of the external electric field inside the material. The two approaches are compared to experimental results. Both the first approach and the experimental results show similar tendencies for optimization of laser ablation in the external electric field.
\end{abstract}


%-----------------------------------------------------------------
\section{Introduction}
Engineering and control of surface and bulk properties of materials using ultrashort laser pulses are highly demanded for modern innovative technologies.
Laser-assisted nanostructuring has already found a lot of applications such as optical waveguides~\cite{davis1996writing}, photonic crystals~\cite{rodenas2019three}, solar fuels, functional colorization of surfaces, friction reduction, controlling of wetting properties~\cite{bonse2016laser,florian2020surface,bonse2020maxwell} and many others. 
However, the quality of the micromachined areas still needs to be improved to fully exploit the potential of laser processing of solids. 
Several methods were proposed to optimize the ablation process, for instance, liquid confinement layers~\cite{zhang2017laser,ivanov2017molecular,shih2020effect} or laser treatment of materials in the presence of different gases~\cite{dominic2021insignificant}. 
An alternative and promising way to control laser micromachining is via application of external magnetic or electric fields~\cite{pandey2015study,singh2016effect,singh2017melt,farrokhi2016magneto,farrokhi2019fundamental,tang2019repulsive,maksimovic2019external,maksimovic2020ablation,zhang2021study,schaefer2021magnetic}. 
An influence of the external magnetic field on the ablation process is still under intense investigations and debates, however, less attention was given so far to the role of an externally applied electric field to the energy deposition and dissipation.
Moreover, the external E-field in the recent investigation~\cite{maksimovic2020ablation} was perpendicular to the laser beam. In contrast, in our experiments, we apply an external electric field along the direction of the laser pulse.

In this paper we report on the influence of an external electric field on the ablation process of a silicon sample following a femtosecond laser pulse. 
Here, we focus on the theoretical description, including the effect of an external electrical field to the density-dependent two temperature model (nTTM)~\cite{van1987kinetics,ramer2014laser,lipp2020solving}.
This model has been successfully applied to simulate laser-matter structuring for different numerical or experimental setups~\cite{lipp2014atomistic,gaudiuso2021laser}. 
The nTTM allows to take into account the transient free-carrier density in contrast to the conventional two-temperature model~\cite{anisimov1974electron,rethfeld2017modelling}, which only tracks carrier and lattice temperature evolution.
We also show results of accompanying experiments, revealing an influence of the external electric field on ablation depth.

The paper is structured as follows: 
In section \ref{sec model} we summarize the main idea, equations and result 
of the nTTM in its original form.
In section \ref{sec efield} we present two different theoretical
approaches aiming to include
the effect of an external electric field. 
Section \ref{sec experiment} is devoted to the experimental part of the project. 
The results of theory and experiment are compared in section \ref{sec comparison}. 
We close with a conclusion and summary. 

%-----------------------------------------------------------------
\section{Density-dependent Two-temperature Model (nTTM)}
\label{sec model}

The excitation of silicon with an ultrashort laser pulse, as well as the following relaxation processes can be described by a density-dependent two-temperature model (nTTM) as introduced by van Driel in Ref.~\cite{van1987kinetics} and presented in the form applied in this work by R\"amer et al.~\cite{ramer2014laser}. 
For completeness, a short description of the model is given here.

The nTTM  tracks the spatially and temporarily resolved evolution of its three main parameters, namely the free-carrier density $n$, electron temperature $T_e$ and the phonon temperature $T_{ph}$. It is governed by the three equations, denoting the energy and particle balance, respectively:

\begin{subequations}
\begin{alignat}{3}
c_{ph} \frac{\partial T_{ph}}{\partial t} &= \bm{\nabla} \cdot (\kappa_{ph} \bm{\nabla} T_{ph}) + g (T_{e} - T_{ph})
\enspace,
\label{eq:u_ph}
\\
c_e \frac{\partial T_{e}}{\partial t} &= (\alpha_\mathrm{SPA} + \alpha_\mathrm{FCA}) I + \beta_\mathrm{TPA} I^2 - \bm{\nabla} \cdot \bm{w} - g (T_{e} - T_{ph}) - \frac{\partial u_e}{\partial n} \frac{\partial n}{\partial t} - \frac{\partial u_e}{\partial \epsilon_g} \frac{\partial \epsilon_g}{\partial t}\
\label{eq:u_e}\enspace,
\\
\frac{\partial n}{\partial t} &= \frac{\alpha_\mathrm{SPA} I}{\hbar \omega_L} + \frac{\beta_\mathrm{TPA} I^2}{2 \hbar \omega_L} + \delta_\mathrm{II} n - \gamma n^3 - \bm{\nabla} \cdot \bm{j} \ \enspace.
\label{eq:n}
\end{alignat}
\end{subequations}

Here, Eq.~(\ref{eq:u_ph}) tracks the evolution of the phonon temperature $T_{ph}$ with the corresponding phonon heat capacity $c_{ph}$. Heat transport in the phonon system is described by the first term on the right-hand side of Eq.~(\ref{eq:u_ph}), where $\kappa_{ph}$ is the phonon heat conductivity. The second term considers electron-phonon coupling, being proportional to 
the temperature difference between electrons and phonons. An important parameter is the electron-phonon coupling parameter $g$.
Here, it is determined as $g=c_e/\tau$, where $c_e$ is the electron heat capacity and $\tau$ is a constant relaxation time.
Equation~(\ref{eq:u_e}) describes the evolution of the electron temperature $T_e$. The first two terms on the right-hand side consider laser absorption with the absorption coefficients for single-photon absorption (SPA) $\alpha_\mathrm{SPA}$, free-carrier absorption (FCA) $\alpha_\mathrm{FCA}$, and two-photon absorption (TPA) $\beta_\mathrm{TPA}$. The third term takes into account heat transport in the electron system by the heat current density $\bm{w}$. 
The fourth term is the electron-phonon coupling, similar as in Eq.~(\ref{eq:u_ph}) and here with the opposite sign to ensure energy conservation. 
The last two terms in Eq.~(\ref{eq:u_e}) are analogous to the left-hand side of the equation. 
All three terms together build the 
total time derivative of the free-carrier energy density $u_e$ 
depending on the electron temperature, free-carrier density and also the band gap energy $\epsilon_g$. The band gap energy is a transient parameter due to its dependence on the free-carrier density and phonon temperature.
The evolution of the free-carrier density $n$ is described by Eq.~(\ref{eq:n}). The first two terms take into account free-carrier generation by SPA and TPA. Here, $\hbar \omega_L$ is the photon energy. The third and fourth term describe impact ionization with the impact ionization coefficient $\delta_\mathrm{II}$ and Auger recombination with the Auger recombination coefficient $\gamma$, respectively. Particle transport with the particle density current $\bm{j}$ is considered by the last term.

For transport in the electron system, electrons and holes are assumed to move together as electron-hole pairs due to the buildup of Dember fields. This is called ambipolar diffusion~\cite{van1987kinetics}. The resulting electron-hole pair density current $\bm{j}$ and electron heat density current $\bm{w}$ are described by

\begin{subequations}
\begin{alignat}{2}
\bm{j} &= \frac{1}{e^2} \frac{\sigma_e \sigma_h}{\sigma_e + \sigma_h} \left[ \bm{\nabla}(\mu_h - \mu_e) + e(S_e - S_h) \bm{\nabla} T_e \right] \enspace,
\label{eq:j}
\\
\bm{w} &= \Pi \bm{j} - (\kappa_e + \kappa_h) \bm{\nabla} T_e \enspace,
\label{eq:w}
\end{alignat}
\end{subequations}

with the electron charge $e$, electric conductivity $\sigma$, chemical potential $\mu$, Seebeck coefficient $S$, Peltier coefficient $\Pi$, and the heat conductivity $\kappa$. The subscripts $e$ and $h$ stand for electrons and holes, respectively.

\begin{figure}[htb]
\centering\includegraphics[width=0.7\linewidth]{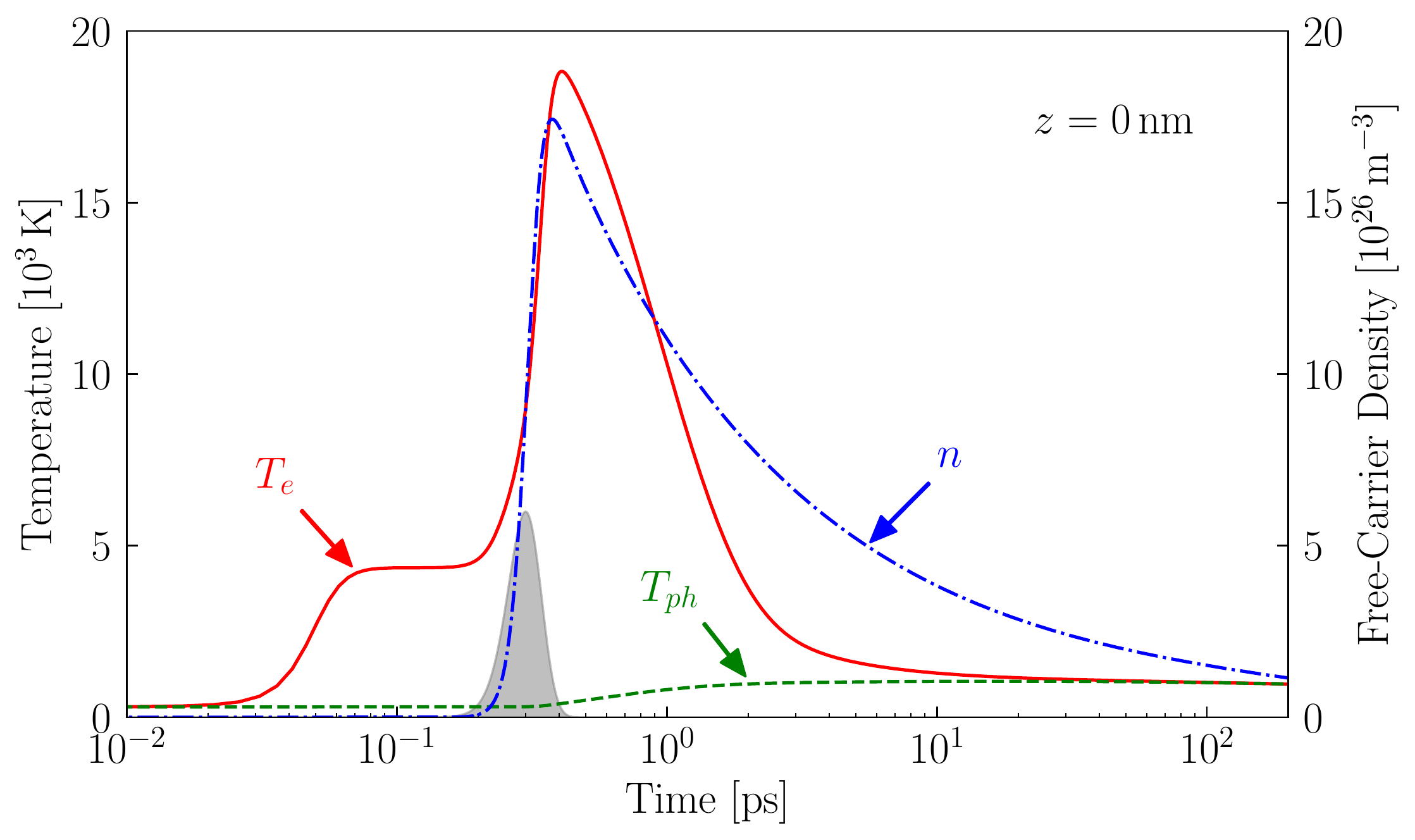}
\caption{Temporal evolution of three main parameters obtained by the nTTM at laser-irradiated surface. The grey shadowed area depicts the laser intensity. For laser parameters, a pulse duration of $\tau=\SI{100}{fs}$, wavelength of ${\lambda=\SI{800}{nm}}$ and a fluence of $F=\SI{130}{mJ/cm^2}$ are used.}
\label{fig:nTTM_temporal}
\end{figure}

The temporal evolution of the three main parameters at the laser-irradiated surface obtained by the nTTM is shown in Fig.~\ref{fig:nTTM_temporal} for a laser pulse duration of $\tau=\SI{100}{fs}$, a laser wavelength of $\lambda=\SI{800}{nm}$ and a sub-threshold laser fluence of $F=\SI{130}{mJ/cm^2}$.

The grey shadowed area depicts the laser intensity. 
At first, the electron system is heated by the far-left flank of the Gaussian laser intensity profile and reaches an electron temperature plateau which corresponds to the kinetic energy of electrons which have been excited across the band gap by SPA. 
As the laser pulse maximum is reached, the carrier-pair density increases rapidly and the electron temperature increases furthermore by FCA, TPA and Auger heating. The latter is caused by an interplay of Auger recombination and the term $- \frac{\partial u_e}{\partial n} \frac{\partial n}{\partial t}$ in Eq.~(\ref{eq:u_e}).
Shortly after the laser intensity peak, the electron temperature and density reach a maximum and decrease again. 
Both quantities are affected by transport processes due to gradients in the chemical potential and temperature. 
Additionally, the decrease in electron temperature is caused by electron-phonon coupling, where the electron and phonon system exchange energy and thereby level their temperatures, and the decrease of carrier density is influenced by Auger recombination.

%-----------------------------------------------------------------
\section{Models for the Effect of an External Electrical Field}
\label{sec efield}

When we apply an electrical field perpendicular to the surface of the irradiated material, electrons and holes will screen the field towards the bulk of the material. Such conditions can be usually described with help of the dielectric function. However, the dielectric function of laser-excited silicon upon and directly after irradiation is changing rapidly, due to the rapidly changing carrier densities and temperatures. Since a full inclusion of transient optical properties is numerically challenging, we study as a first step, two extreme cases of the microscopic response of the charges within the bulk material. 
The first case maintains the assumption of ambipolar diffusion within the material: An external electrical field is capable to accelerate the mobile electrons within the 
skin layer. Further inside the material, the external electrical field is screened and the accelerated electrons move freely through the bulk. With the concept of ambipolar diffusion, which essentially avoids the buildup of further internal electrical fields, electrons and holes will then move together through the bulk. Within this approach, we do not calculate the details of the charge-acceleration within the skin layer, but assume a constant drift velocity for electron-hole pairs. We describe the extension of our model and the obtained results in subsection \ref{sub sec current}. The second case studies the other extreme, which is the full charge separation of electrons and holes. Depending on the mobility of the charge carriers, they are able to annihilate the electrical field within the material. We describe this approach and the results in subsection \ref{sub sec charge}.

%-----------------------------------------------------------------
\subsection{Additional Carrier-pair Current}
\label{sub sec current}

Here, we study the influence of a constant drift velocity of the electron-hole pairs on the temporal and spatial free-carrier density as well as on the electron and phonon temperature.

\begin{figure}[H]
\centering\includegraphics[width=7cm]{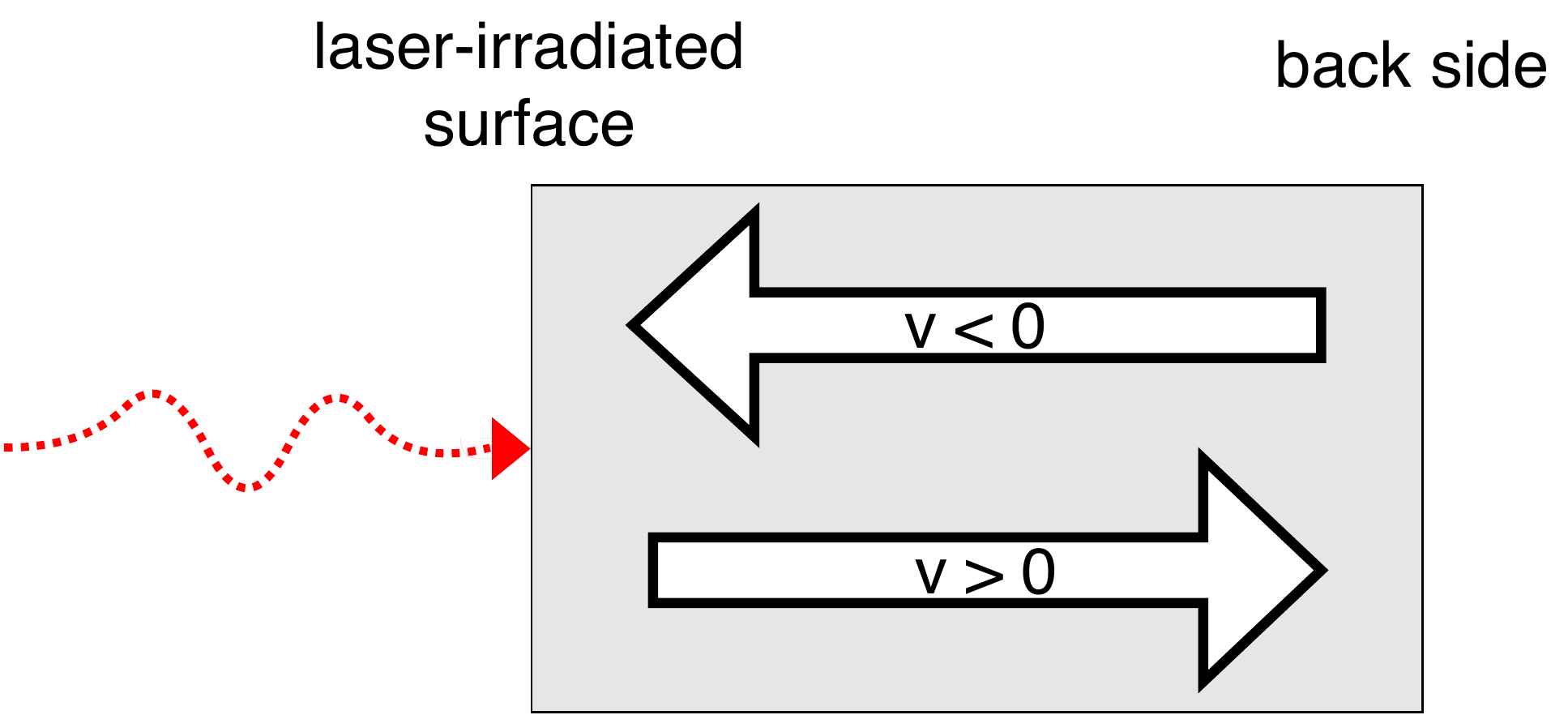}
\caption{Schematic view of laser-irradiated material. The arrows indicate the direction of the additional current $\bm{j_\mathrm{add}}$ for a negative and positive drift velocity $v$.}
\label{fig:convention}
\end{figure}

A current of 
\begin{equation}
\bm{j_\mathrm{add}} = v \cdot n
\label{eq:jadd}
\end{equation}
is added to the electron-hole pair density current $\bm{j}$ in Eq.~(\ref{eq:j}).
Here, $v$ is a constant drift velocity and $n$ is the density of electron-hole pairs
as above. The direction of the drift velocity is 
perpendicular to the laser-irradiated surface, with the sign given according to Figure~\ref{fig:convention}.

Figure~\ref{fig:convention} depicts a schematic view of the laser-irradiated material. The arrows indicate the direction of the additional carrier-pair current $\bm{j_\mathrm{add}}$ for different signs of the drift velocity $v$. For a negative sign of $v$, the current is headed from the back side of the material towards the laser-irradiated surface, for a positive sign in the opposite direction.

\begin{figure}[H]
\centering\includegraphics[width=0.85\textwidth]{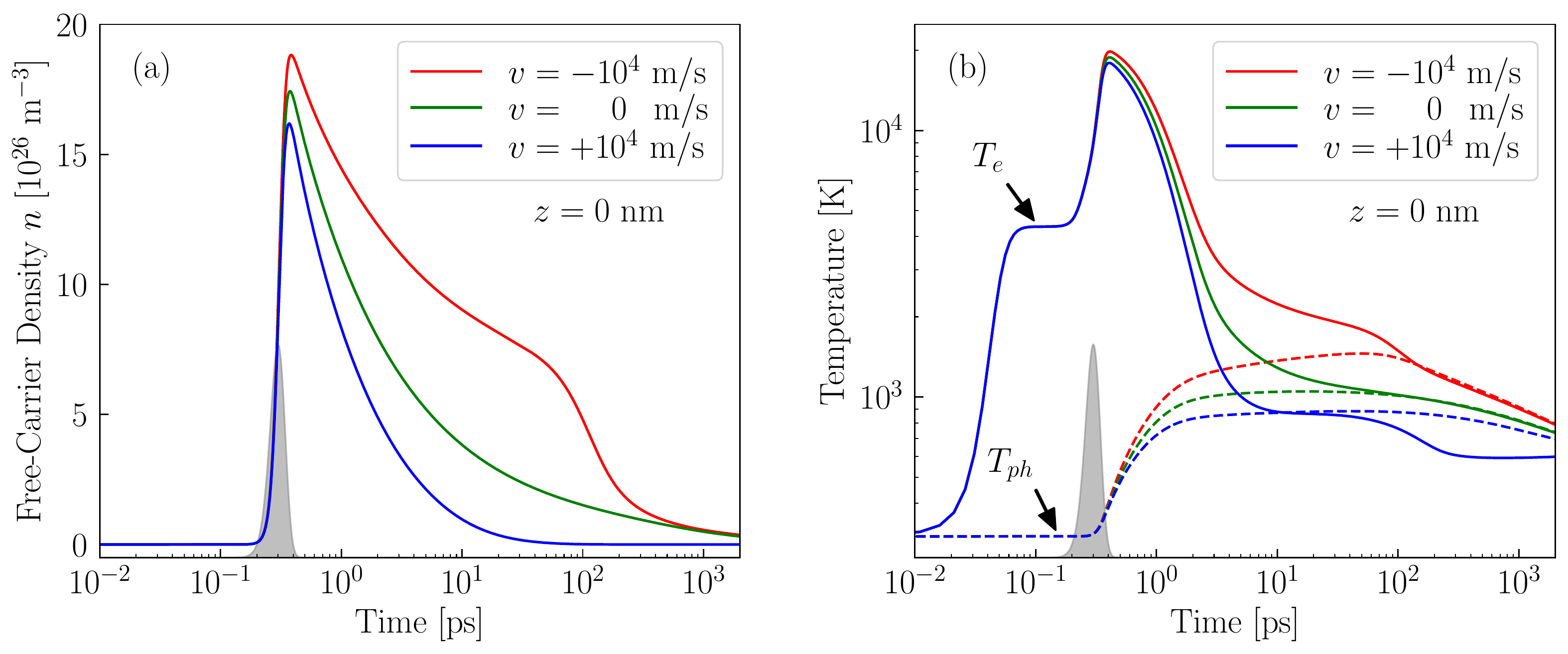}
\caption{Temporal evolution of the free-carrier density $n$ (a) and the electron temperature $T_e$ and phonon temperature $T_{ph}$ (b) at the laser-irradiated surface for different directions of the drift velocity $v$.}
\label{fig:addcurrent_temporal}
\end{figure}

Figure~\ref{fig:addcurrent_temporal} shows the temporal evolution of the three main parameters at the laser-irradiated surface for different directions of the additional current $\bm{j_\mathrm{add}}$. A laser wavelength of $\lambda = \SI{800}{nm}$, pulse length of $\tau=\SI{100}{fs}$ and a fluence of $F=\SI{130}{mJ/cm^2}$ is used. For the negative (positive) direction, the free-carrier density $n$ and carrier temperature $T_e$ are increased (decreased) by the additional current. As a consequence, the phonon temperature $T_{ph}$ is also increased (decreased).

\begin{figure}[H]
\centering\includegraphics[width=0.49\textwidth]{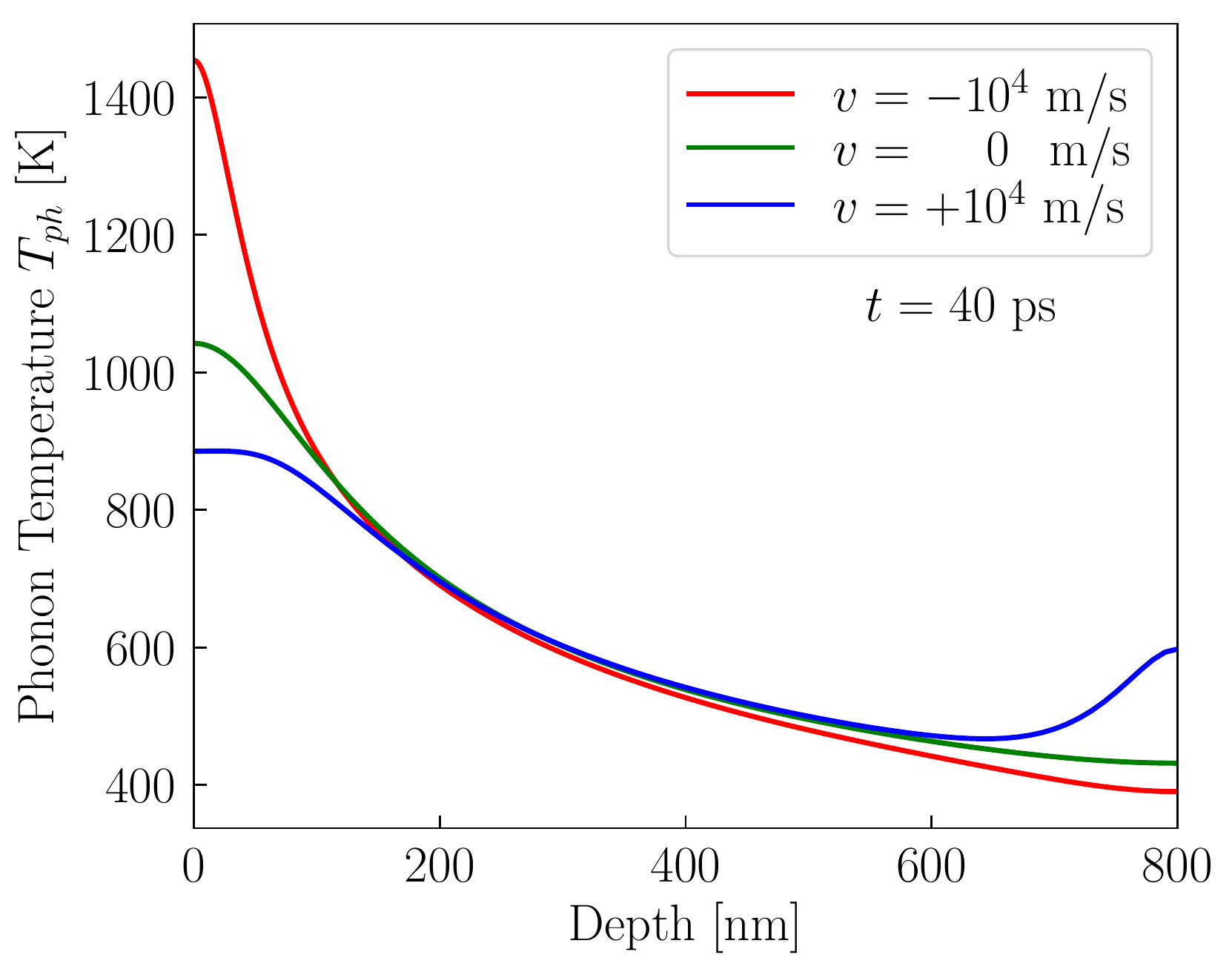}
\caption{Spatial phonon temperature profile for different drift velocities $v$, $40 \, \text{ps}$ after laser irradiation.}
\label{fig:addcurrent_spatial}
\end{figure}

Figure~\ref{fig:addcurrent_spatial} depicts the spatial phonon temperature profile within the sample of $\SI{800}{nm}$ thickness. At the chosen time of $\SI{40}{ps}$, the phonon temperature near the surface has roughly reached its maximum. For the negative direction of the additional current, i.e. pointing towards the surface, a steeper phonon temperature profile is reached near the laser-irradiated surface. In contrast to that, the profiles are flattened for the positive direction. For this direction, pointing towards the bulk, a considerable increase of phonon temperature can be observed at the back side of the material.

%-----------------------------------------------------------------
\subsection{Charge Separation}
\label{sub sec charge}

In this approach, free electrons and holes are treated separately to account for the charge separation in an external electric field. Thus, the free-carrier density $n$ is replaced by the free electron density $n_e$ and the hole density $n_h$. However, free electrons and holes are still assumed to posses a common temperature $T_e$.

Therefore, the evolution of the free-carrier density in Eq.~(\ref{eq:n}) is modified to account for separate densities

\begin{equation}
\frac{\partial n_{e,h}}{\partial t} = \frac{\alpha_\mathrm{SPA} I}{\hbar \omega_L}+ \frac{\beta_\mathrm{TPA} I^2}{2 \hbar \omega_L} + \delta_{eeh} n_e + \delta_{ehh} n_h - \gamma_{eeh} n_e^2 n_h - \gamma_{ehh} n_e n_h^2 - \bm{\nabla} \cdot \bm{j_{e,h}} \, ,
\label{eq:n_separate}
\end{equation}

where the subscripts $e$ and $h$ represent electrons and holes, respectively. The first two terms in Eq.~(\ref{eq:n_separate}) which describe laser excitation are the same for both carrier types since electrons and holes are created as pairs. The next four terms account for impact ionization and Auger recombination. Both events which involve two electrons and one hole ($eeh$) and events which involve one electron and two holes ($ehh$) are considered. The last term describes particle transport with two separate particle density currents which read

\begin{equation}
\bm{j_{e,h}} = \mp \frac{\sigma_{e,h}}{e} \left[ \bm{E} + \frac{\bm{\nabla} \mu_{e,h}}{e} +  S_{e,h} \bm{\nabla} T_{e} \right] \ .
\label{eq:j_separate}
\end{equation}

Here, $\bm{E}$ is the internal electric field which is found by solving $\bm{\nabla} \cdot (\epsilon \bm{E}) = e(n_h - n_e)$ with the boundary condition, that the electric field at the laser-irradiated surface is  $\bm{E}(z=0) = \bm{E}_{ext}/\epsilon$. Here, $\bm{E}_{ext}$ is the applied external electric field and $\epsilon$ the static dielectric constant.

Furthermore, the heat current of electrons and holes, respectively, read

\begin{equation}
\bm{w_{e,h}} = \Pi_{e,h} \, \bm{j_{e,h}} - \kappa_{e,h} \bm{\nabla} T_e \, ,
\label{eq:w_separate}
\end{equation}

with the total carrier heat current given by $\bm{w} = \bm{w_e} + \bm{w_h}$ .
The material properties, newly used in this extension of the model, are listed in Table~\ref{table:charge-sep}. The mobilities of electrons and holes, respectively, are both assumed to be constant.

\begin{table}[H]
\caption{Applied material properties for the 'Charge Separation' model. All other parameters are identical to the ones used in the nTTM~\cite{ramer2014laser}.}
\label{table:charge-sep}
\centering
\begin{centering}
 \begin{tabular}{p{7cm}  p{8cm}} 
\\[3pt]
 \hline
 Quantity & Value \\
 \hline\\

\multicolumn{2}{c}{See Ref.~\cite{van1987kinetics} for the calculation of the transport coefficients $\sigma_{e,h}$, $S_{e,h}$ and $\Pi_{e,h}$.}\\[5pt]

\multicolumn{2}{c}{Carrier properties}\\[1pt]
Electron mobility $\varsigma_e$ & $\SI{0.0085}{m^2 / V s}$ \, \cite{lipp2020solving} \\[0.5pt]
Hole mobility $\varsigma_h$ & $\SI{0.0019}{m^2 / V s}$ \, \cite{lipp2020solving} \\[0.5pt]
Auger coefficient (eeh) $\gamma_{eeh}$ & $\SI{2.8e-31}{cm^6/s}$ \, \cite{dziewior1977auger} \\[0.5pt]
Auger coefficient (ehh) $\gamma_{ehh}$ & $\SI{0.99e-31}{cm^6/s}$ \, \cite{dziewior1977auger} \\[0.5pt]
Impact ionization rate $\delta_\mathrm{eeh} = \delta_\mathrm{ehh} =  \frac{\delta_{II}}{2}$ & $\frac{1}{2} \times 3.6 \times 10^{10} \exp [-1.5 \, E_\mathrm{gap} / (k_B T_e)] \, \mathrm{s}^{-1}$ \, \cite{van1987kinetics} \\[10pt]

\multicolumn{2}{c}{Optical properties}\\[1pt]
Static dielectric constant $\epsilon$ & $11.97$ \, \cite{samara1983temperature}  \\[10pt]

\hline
\hline

\end{tabular}
\end{centering}

\end{table}

\begin{figure}[H]
\centering\includegraphics[width=0.6\textwidth]{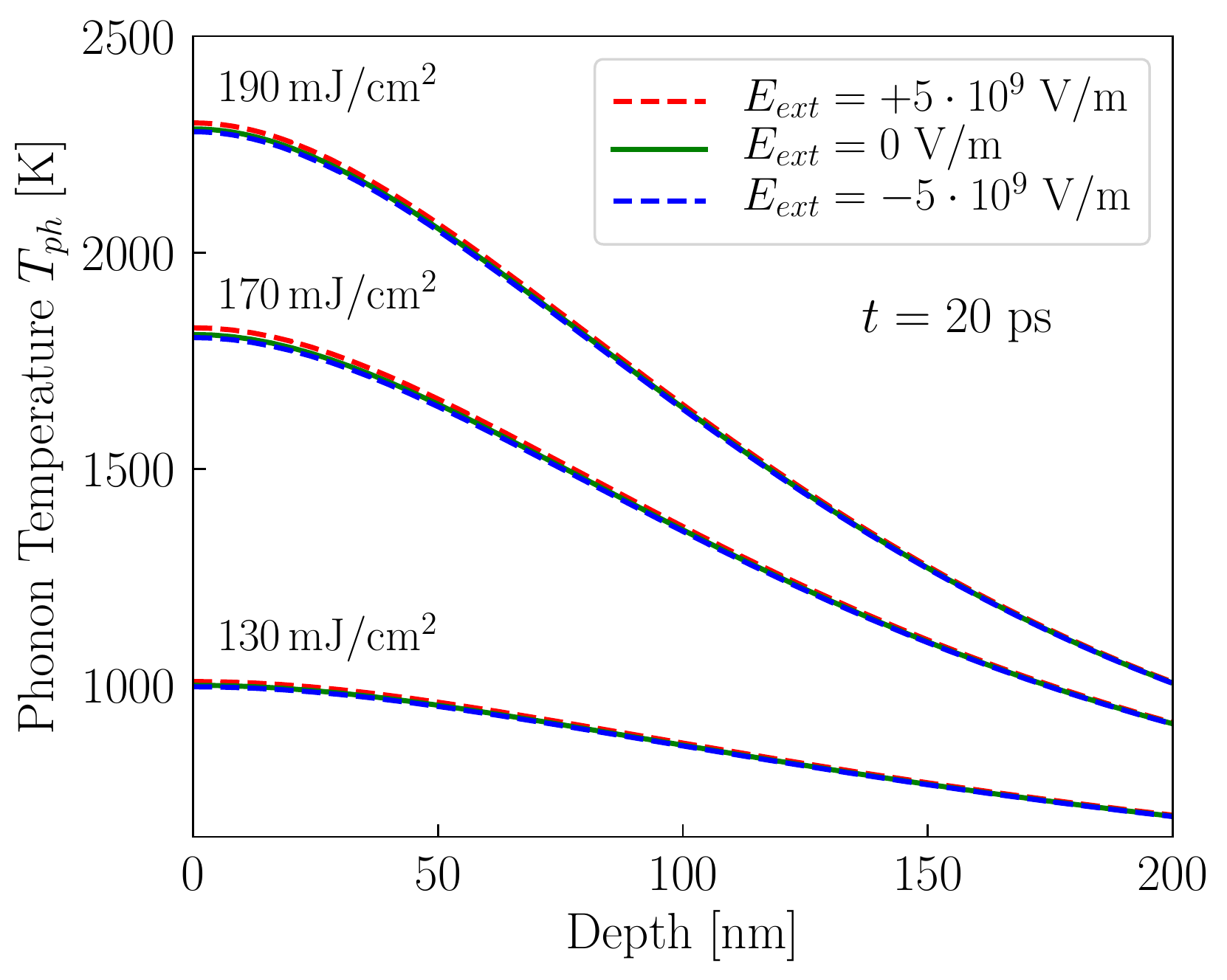}
\caption{Spatial phonon temperature profile after $20 \, \text{ps}$ for different external electric fields and fluences.}
\label{fig:efield_phonon}
\end{figure}

 Fig.~\ref{fig:efield_phonon} shows the spatial profile of the lattice temperature after $\SI{20}{ps}$ obtained with this model for different fluences and different external electric fields. As in subsection  \ref{sub sec current}, a laser wavelength of $\lambda=\SI{800}{nm}$ and a pulse length of $\tau=\SI{100}{fs}$ have been applied. 
In all investigated cases, the effect of the electric field on the phonon temperature profile is small, despite of large electric field strengths.

\begin{figure}[H]
\centering\includegraphics[width=0.85\textwidth]{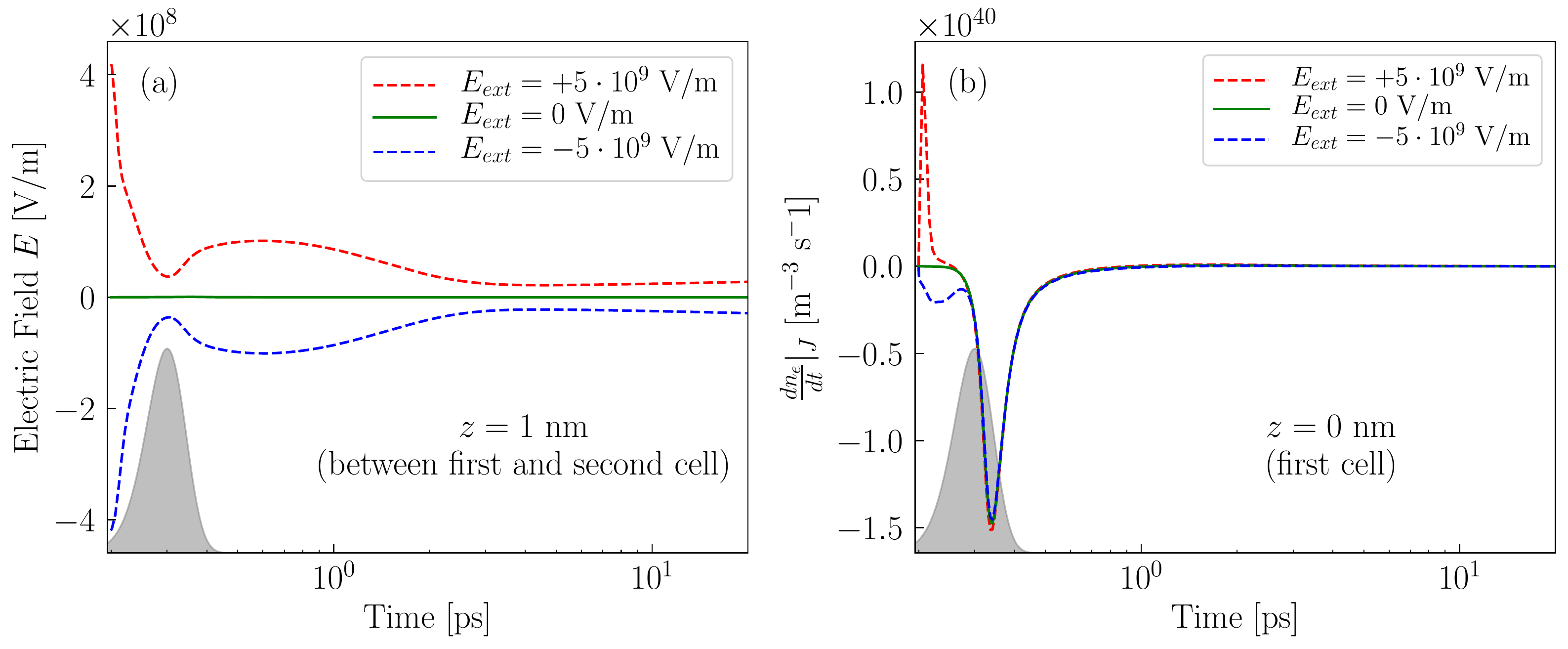}
\caption{Temporal evolution of the internal electric field $E$ (a) and gain in electron density by transport $\frac{dn_e}{dt}|_{\bm{J}}$ (b).}
\label{fig:efield}
\end{figure}

Figure~\ref{fig:efield} shows the temporal evolution of the electric field between the first and second discretization cell at a depth of $\SI{1}{nm}$ (a), as well as the change in electron density by transport in the first cell $\frac{dn_e}{dt}|_{\bm{J}}$ (b) for a laser fluence of $F=\SI{190}{mJ/cm^2}$. The electric field strength decreases significantly before the laser pulse maximum is reached. This is caused by the rapid change in electron density, which corresponds to the charge separation of the electron-hole pairs which are generated by the left flank of the Gaussian laser pulse. After the laser pulse maximum, the electric field strength increases again due to particle currents caused by the relaxation of gradients in the material. The increase in electric field strength has no significant effect on the change in electron density after the laser pulse maximum, compared to the case of no applied external electric field.

%-----------------------------------------------------------------
\subsection{Comparison of the models}
\label{sub sec compare}

For the two presented models, very different assumptions were made which also yield very different results. For the model presented in section~\ref{sub sec current}, electrons and holes are assumed to move as pairs such that quasi neutrality is not broken and a significant effect on the phonon temperature is observed for a constant drift velocity. Contrary to that, electrons and holes are separated by the external electric field in the model presented in section~\ref{sub sec charge}. This charge separation leads to rapid shielding inside the material and thus the effect of the external electric field 
on the phonon temperature is marginal.
Note that further studies are planned, particularly considering the microscopic development of a free particle drift on one hand and a transient dielectric function on the other hand. Therefore, we cannot compare the results quantitiatively. However, we have performed experimental investigations to estimate the quantity of the effect of an 
external electric field. They are presented in the next section.

%%%%%%%%%%%%%%%%%%%%-----------------------------------------------------------------
\section{Experiment}
\label{sec experiment}

\begin{figure}[H]
\centering\includegraphics[width=0.49\textwidth]{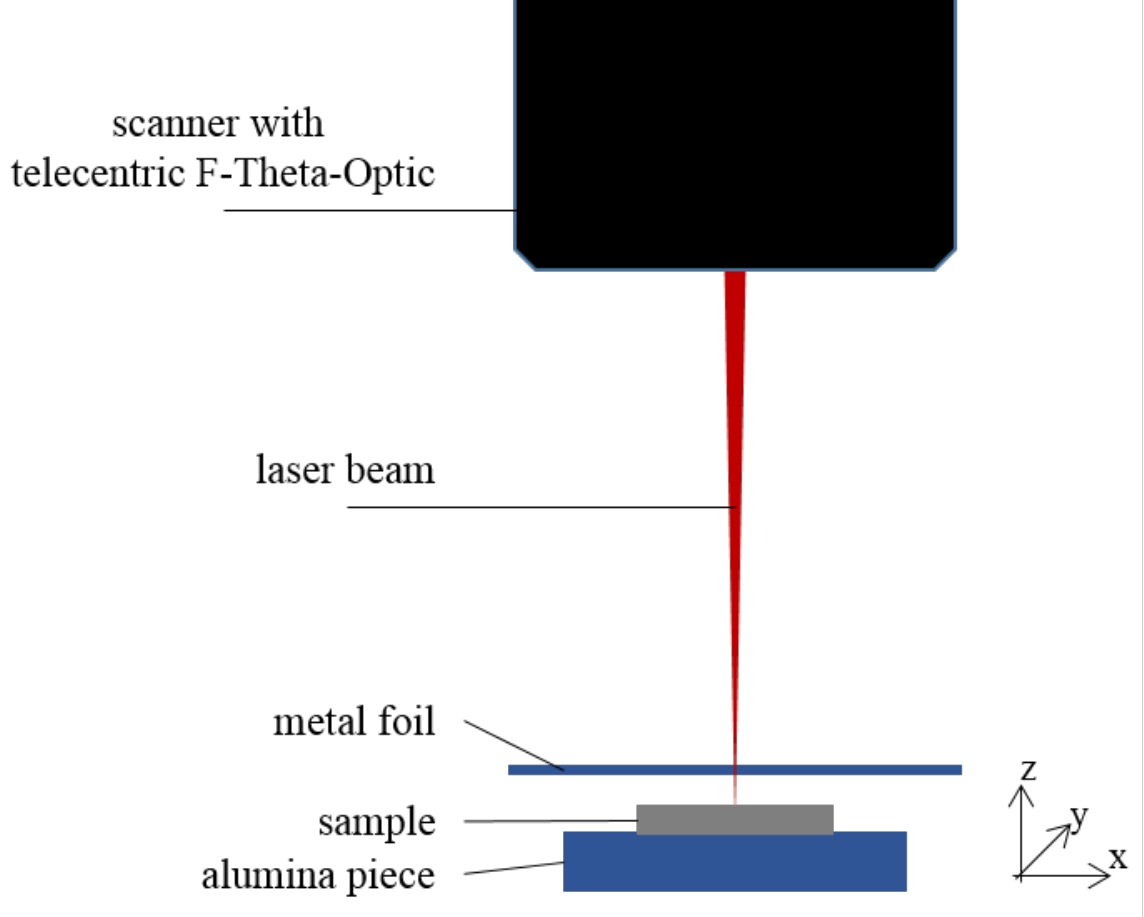}
\caption{Schematic view of the experimental setup.}
\label{fig:experiment-setup}
\end{figure}

Figure~\ref{fig:experiment-setup} shows the schematic diagram of the experimental setup. For the experiments, a femtosecond fiber laser system (BlueCut, Menlo Systems GmbH), with a pulse length of about $\SI{400}{fs}$, a central wavelength of $\SI{1030}{nm}$, and maximum pulse energy of $\SI{10}{\mu J}$, was used to irradiate the surface. The laser polarization was linear. To focus the laser beam onto the sample surface a telecentric F-Theta lens with $f_l=\SI{100}{mm}$ focal length, which provides a beam waist of $w_0=\SI{9.5}{\mu m}$, was applied.

The electric field was generated parallel to the laser beam. To achieve this the sample is positioned between a metallic foil and a metallic sample, which were attached to a voltage generator. The top electrode is a $\SI{100}{\mu m}$ thick metal foil and to avoid breaking through in silicon when a high electric field is applied it is placed $\SI{1.5}{mm}$ above the sample surface. Additionally, to guarantee a nearly homogenous electric field within the processing area a through-hole with $\SI{1}{mm}$ in diameter was processed in the middle of the top electrode. The electric field strength was set to be $\SI{411.5}{V/mm}$ and we investigated the influence in both directions. The electric field direction is regarded as plus ('$+$') direction when it points from the top to the bottom and vice versa for the negative ('$-$') direction.In this study, single-pulse laser ablation on the polished side of the silicon pieces with the dimensions of $10\times10\times0.5 \, \text{mm}^3$ was performed across a range of laser pulse energies. The ablated profile was measured with a confocal microscope (Zeiss smart proof 5, Carl Zeiss AG). 

\begin{figure}[H]
\centering\includegraphics[width=0.49\textwidth]{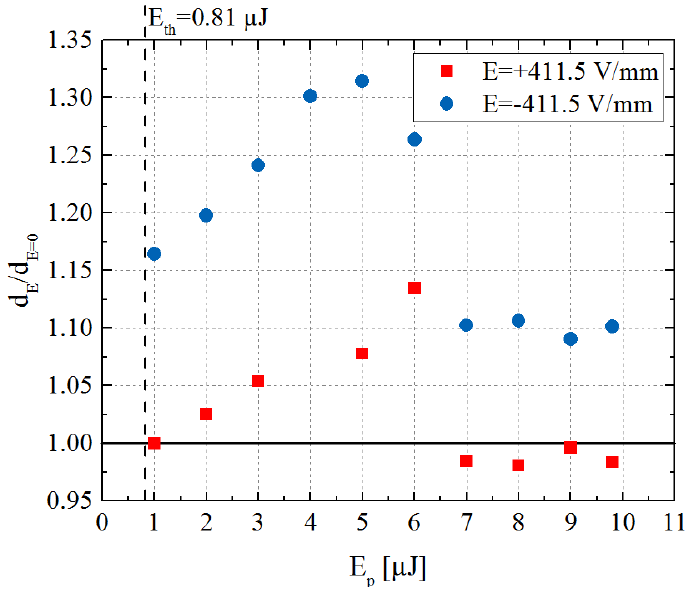}
\caption{Change in removal depth in dependence on the applied pulse energy $E_p$.}
\label{fig:single-pulse-ablation}
\end{figure}

Figure~\ref{fig:single-pulse-ablation} shows how the ablation depth change in the presence of an electric field in dependence on the applied pulse energy $E_p$. 
The change of ablation depth will be considered by the ratio of the ablation depth with an electric field ($d_E$) to without electric field ($d_{E=0}$). 
With increasing pulse energy the holes become deeper. 
However, at a pulse energy of $\sim \SI{6}{\mu J}$, a drop in the deepness of the holes occurs. 
By an applied positive electric field the holes become even shallower compared to the ablated structures without an applied electric field. 
Moreover, for high pulse energies, the change in removal depth remains almost constant. 
In general, the investigations show that the most pronounced effect is present by applying a negative electric field during the laser ablation depth. The created holes are always approximately $10\%$ to $20\%$ deeper compared to an applied positive electric field.

%%%%%%%%%%%%%%%%%%%%-----------------------------------------------------------------
\section{Comparison}
\label{sec comparison}

The experiments reveal a significant effect of the external electric field on ablation depth, while the 'Charge Separation' model described in section~\ref{sub sec charge} only shows marginal effects on phonon temperature for larger electric field strengths. However, a qualitative comparison can be made between the experiments and the 'Additional Carrier-pair Current' model described in section~\ref{sub sec current}.

\begin{figure}[htb]
\centering\includegraphics[width=0.6\textwidth]{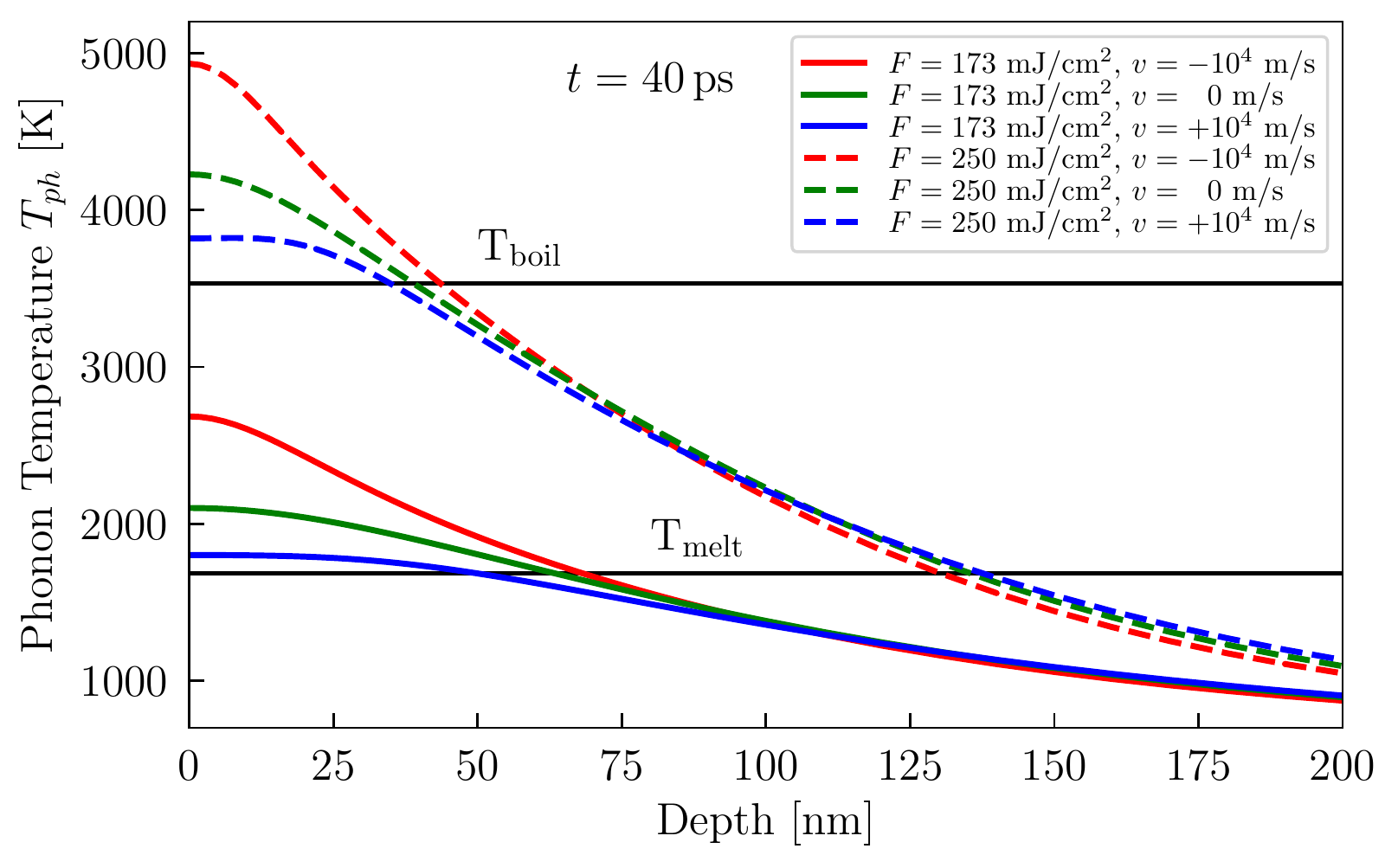}
\caption{Spatial phonon temperature profile obtained by the 'Additional Carrier-pair Current' model described in section~\ref{sub sec current}, $\SI{40}{ps}$ after laser irradiation for different laser fluences.}
\label{fig:addcurrent_fluence}
\end{figure}

Fig.~\ref{fig:addcurrent_fluence} shows the spatial phonon temperature profile near the laser-irradiated surface obtained by the 'Additional Carrier-pair Current' model.
A laser wavelength of $\lambda=\SI{800}{nm}$, pulse length of $\tau=\SI{100}{fs}$ and two different fluences were used. 
The lower fluence of $F=\SI{173}{mJ/cm^2}$ lies slightly above the melting threshold and the higher fluence of $F=\SI{250}{mJ/cm^2}$ is chosen, such that phonon temperatures near the laser-irradiated surface above the boiling temperature are reached. 
The results show a reversal of the effect of the additional current on melting depth for higher fluences. 
For the lower fluence, the melting depth is increased for the negative direction and decreased for the positive direction. The opposite effect is observed for the higher fluence.

Since the electron mobility is higher than the hole mobility, we attribute the positive drift direction in this model to the negative field direction in the experiment. The higher melting depth for the positive drift direction at the higher fluence corresponds to the observation of higher ablation depths for negative field direction in the experiment, where laser fluences far above the ablation threshold were used. From these results we conclude that a shallower phonon temperature profile lowers heat loss near the laser-irradiated surface and thus is advantageous to achieve higher ablation depths.

%%%%%%%%%%%%%%%%%%%%-----------------------------------------------------------------   
\section{Summary and Conclusion}

We presented two approaches on how to describe the effects of an external electric field on the initial stage of laser ablation in silicon. Also experimental measurements have been performed, which show an effect of the external electric field on ablation depth. Similar tendencies are observed for the first approach, in which ambipolar diffusion is assumed and a carrier-pair current with constant drift velocity is added. The second approach which accounts for charge separation and shielding of the external electric field inside the material shows only marginal effects on the spatial lattice temperature profile. Further studies are needed to refine both approaches and to allow for quantitative comparisons between theory and experiment.

\subsection*{Disclosures}

The authors declare that there is no conflict of interest.

\subsection*{Acknowledgement}

We thank the Federal Ministry of Education and Research (BMBF), project ''AssistAb'', FKZ 13N14867 and 13N14868.

\bibliographystyle{unsrt}

\end{document}